\documentclass[aps,prd,superscriptaddress,long,notitlepage,balancelastpage,nofootinbib,floatfix,twocolumn]{revtex4-2}
\pdfoutput=1
\usepackage{amsmath,mathtools,amssymb,amsthm,amsxtra,overpic,bbm,epsfig,subfigure,url,bm}
\usepackage{mathrsfs}
\usepackage{color,xcolor}
\usepackage{comment}
\usepackage{float}
\usepackage{enumitem}
\usepackage{slashed}
\usepackage{multirow}
\usepackage{appendix}
\usepackage[colorlinks=true,linkcolor=blue,citecolor=blue]{hyperref}

\begin{document}
\title{ Explanation of the 95 GeV $\gamma\gamma$ and $b\bar{b}$ excesses \\
in the Minimal Left-Right Symmetric Model}

\author{\bf P. S. Bhupal Dev}
\email{bdev@wustl.edu}
\affiliation{Department of Physics and McDonnell Center for the Space Sciences, Washington University, St.~Louis, Missouri 63130, USA}

\author{\bf Rabindra N. Mohapatra}
\email{rmohapat@umd.edu}
\affiliation{Maryland Center for Fundamental Physics and Department of Physics, University of Maryland, College Park, Maryland 20742, USA}

\author{\bf Yongchao Zhang}
\email{zhangyongchao@seu.edu.cn}
\affiliation{School of Physics, Southeast University, Nanjing 211189, China}

\date{\today}

\begin{abstract} 
We propose a simple interpretation of the $\gamma\gamma$ excesses reported by both CMS and ATLAS groups at 95 GeV together with the LEP excess in the $Zb\bar{b}$ channel around the same mass in terms of a neutral scalar field in the minimal left-right symmetric model (LRSM). We point out that the scalar field which implements the seesaw mechanism for neutrino masses has all the right properties to explain these observations, without introducing any extra fields. The key point is that this scalar particle is hardly constrained because it couples only to heavy right-handed particles. As a result, the diphoton decay mode receives contributions from both mixing with the Standard Model (SM) Higgs and the heavy charged bosons in the LRSM. The latter depends on the $SU(2)_R\times U(1)_{B-L}$ symmetry breaking scale $v_R$. The complete allowed parameter space for explaining the 95 GeV excesses in this model can be probed with the high-precision measurements of the SM Higgs mixing with other scalars at the high-luminosity LHC and future Higgs factories. 
\end{abstract}

\maketitle

\section{Introduction}  
Since the landmark discovery of the 125 GeV Higgs boson~\cite{ATLAS:2012yve, CMS:2012qbp}, its precision studies, as well as searches for additional Higgs bosons, are among the top priorities at the LHC~\cite{Dawson:2022zbb}. Along with the heavy Higgs searches, the mass range below 125 GeV is also receiving increased scrutiny, to make sure that no stone remains unturned. In this context, it is intriguing to note that the CMS Collaboration has recently reported  a 2.9$\sigma$ global ($1.9\sigma$ local) excess  in the diphoton channel with an invariant mass of 95.4 GeV~\cite{CMS:2023yay}. A similar analysis by the ATLAS Collaboration has found a 1.7$\sigma$ local excess at the same mass~\cite{ATLAS-CONF-2023-035}. In addition, CMS has reported another excess in the $\tau^+\tau^-$ channel around 100 GeV with $2.6\sigma$ local ($2.3\sigma$ global) significance~\cite{CMS:2022goy}. A previously seen excess in the LEP data in the $e^+e^-\to Z(H\to b\bar{b})$ channel at 2.3$\sigma$ local significance around the same mass~\cite{LEPWorkingGroupforHiggsbosonsearches:2003ing} has added further supporting curiosity regarding these hints from the LHC, setting off a deluge of theoretical activity~\cite{Cao:2016uwt,Fox:2017uwr,Haisch:2017gql,Biekotter:2017xmf,Liu:2018xsw,Biekotter:2019kde,Cline:2019okt, 
Choi:2019yrv,Kundu:2019nqo,Sachdeva:2019hvk,Cao:2019ofo,Aguilar-Saavedra:2020wrj, Abdelalim:2020xfk,Biekotter:2021qbc,Heinemeyer:2021msz,Biekotter:2022jyr,Benbrik:2022azi,Iguro:2022dok,Li:2022etb,Banik:2023ecr,Biekotter:2023jld,Bonilla:2023wok,Azevedo:2023zkg,Escribano:2023hxj,Biekotter:2023oen,Belyaev:2023xnv,Ashanujjaman:2023etj,Bhattacharya:2023lmu,Aguilar-Saavedra:2023tql,Dutta:2023cig,Ellwanger:2023zjc,Cao:2023gkc,Borah:2023hqw,Ahriche:2023hho,Arcadi:2023smv,Mulaudzi:2023khg, Ahriche:2023wkj, Chen:2023bqr}.\footnote{Most of these papers focus only on one or two of these three excesses ($\gamma\gamma$, $\tau^+\tau^-$ and $b\bar{b}$). A simultaneous explanation of all three excesses in a single model remains a challenging task.} 
Much of the  excitement comes from the fact that the Standard Model (SM) is definitely an incomplete theory of Nature. Therefore, any statistically significant deviation from the SM expectation could be a sign of the long-awaited new physics, which makes `anomaly chasing' a worthwhile exercise~\cite{Fischer:2021sqw, Crivellin:2023zui}. Specifically, if these hints for a new 95 GeV scalar are confirmed in the future at 5$\sigma$ confidence, it will be an indication that in the very least the scalar sector of the SM needs an extension. The existence of neutrino mass, evidence for dark matter and the observed matter-antimatter asymmetry of the universe in any case provide more compelling motivations for the existence of beyond SM physics, which requires an extension of the SM either in the scalar, gauge, or fermion sector.  

The minimal left-right symmetric model (LRSM) extension of the SM, based on the gauge group $SU(2)_L\times SU(2)_R\times U(1)_{B-L}$~\cite{Pati:1974yy, Mohapatra:1974gc,Senjanovic:1975rk}, provides a solution to several of the abovementioned problems of the SM, such as the nonzero neutrino masses via the seesaw mechanism~\cite{Minkowski:1977sc,Mohapatra:1979ia, Yanagida:1979as, Gell-Mann:1979vob, Glashow:1979nm,Mohapatra:1980yp}, origin of matter-antimatter asymmetry via leptogenesis~\cite{Fukugita:1986hr, Frere:2008ct, Dev:2014hro, Dev:2015khe, Dhuria:2015cfa, Dev:2019ljp} and a warm dark matter candidate in terms of the lightest right-handed neutrino~\cite{Nemevsek:2012cd,Nemevsek:2023yjl}. It furthermore allows for grand unification of gauge couplings to an $SO(10)$ group~\cite{Chang:1984qr} with a low $W_R$ scale. In this paper we point out that this well-motivated model contains a real neutral scalar boson (denoted below by $H_3^0$), as an intrinsic part of the theory, with the right properties to explain the 95 GeV excesses in both the LHC and LEP data. This field is the real component of the $SU(2)_R$ scalar triplet field with $B-L=2$ that implements the seesaw mechanism for neutrino masses~\cite{Mohapatra:1979ia, Mohapatra:1980yp}. In contrast, in many extensions of the SM considered to date for the purpose of the 95 GeV anomaly, one adds extra Higgs doublets along with singlets, or considers supersymmetric extensions which often need extra scalar fields just to accommodate the excess. The interesting point about this scalar field is that it only couples to the heavy right-handed particles.  As a result its mixing with the SM Higgs boson plays the main role in its production and decay. 
In addition, in the LRSM the $H_3^0\to \gamma\gamma$ decay channel explicitly depends on the $SU(2)_R$-breaking scale $v_R$, as already noticed in Refs.~\cite{Dev:2016dja,Dev:2016nfr,Dev:2017dui}. Thus, the future tests of these excesses, in particular in the diphoton channel, could at least provide some information on the $v_R$ scale for parity restoration.

The rest of the paper is organized as follows: in Section~\ref{sec:model} we give a brief overview of the salient features of the LRSM. In Section~\ref{sec:excess} we discuss the identification of the $SU(2)_R$-breaking scalar $H^0_3$ as the $95$ GeV excess hinted at LHC and LEP. Some representative benchmark scenarios are provided in this section, including the cases without and with tree-level flavor-changing neutral current (FCNC) couplings of $H_3^0$ with the SM fermions. Section~\ref{sec:implications} is devoted to study of the implications of this identification such as the decay of $H_3^0$ and the FCNC constraints, as well as some future collider tests of the model. In Section~\ref{sec:conclusion} we conclude with a summary of the main results. The calculational details for the rare decay $h \to H_3^0 b\bar{b}$ in the LRSM is presented in Appendix~\ref{sec:appendix}. 

\section{The model} 
\label{sec:model}

Let us start with a brief review of the minimal LRSM for neutrino masses~\cite{Pati:1974yy,Mohapatra:1974gc,Senjanovic:1975rk, Mohapatra:1979ia}. The model is based on the gauge group $SU(3)_c \times SU(2)_L\times SU(2)_R\times U(1)_{B-L}$, with 
the quantum numbers of the fermion and scalar fields under the gauge group given in Table~\ref{tab:quantumnumbers}. We omit the $\Delta_L$ $(1,\, 3,\, 1,\, +2)$ field in what follows, since we will deal with the effective field theory at lower energies below the $D$-parity breaking scale and the $\Delta_L$ triplet is expected to remain at the higher scale in this version~\cite{Chang:1983fu}. This also guarantees that the neutrino masses are given by the simple type-I seesaw formula~\cite{Mohapatra:1979ia}, linking the neutrino masses to the $W_R$ boson mass scale. This version is also more amenable to $SO(10)$ embedding with a TeV-scale $W_R$ boson~\cite{Chang:1984qr}.

\begin{table}[t!]
\centering
\caption{Gauge quantum numbers of the fermion and scalar fields in the minimal LRSM under the gauge group $SU(3)_c\times SU(2)_L\times SU(2)_R\times U(1)_{B-L}$.}
\label{tab:quantumnumbers}
\begin{tabular}{ccc}
\hline\hline
\multicolumn{2}{c}{Particles} & Quantum numbers \\ \hline
%{\rm Fermion sector}&\\\hline
\multirow{4}{*}{fermions} & $Q_L = \left(\begin{array}{c}u_L\\d_L\end{array}\right)$ & $(3, 2, 1, 1/3)$ \\
& $Q_R = \left(\begin{array}{c}u_R\\d_R\end{array}\right)$ & $(3, 1, 2, 1/3)$ \\
& $\ell_L = \left(\begin{array}{c}\nu \\ e_L\end{array}\right)$ & $(1, 2, 1, -1)$ \\
& $\ell_R = \left(\begin{array}{c} N \\ e_R\end{array}\right)$ & $(1, 1, 2, -1)$ \\ \hline
%{\rm Scalar sector} &\\\hline
\multirow{2}{*}{scalars} & $\Phi = \left( \begin{array}{cc} \phi_1^0 & \phi_2^+  \\ \phi_1^- & \phi_2^0 \end{array} \right)$ & $(1,2,2,0)$ \\
%& $\Delta_L$ & $(1,3, 1,+2)$\\
& $\Delta_R = \left( \begin{array}{cc} \Delta_R^+/\sqrt2 & \Delta_R^{++}  \\ \Delta_R^0 & -\Delta_R^+/\sqrt2 \end{array} \right)$ & $(1,1, 3,+2)$ \\
\hline\hline
\end{tabular}
\end{table}

The Yukawa Lagrangian for the model  is given by
\begin{eqnarray}
{\cal L}_Y &=&
h_q\bar{Q}_L \Phi Q_R + 
h'_q\bar{Q}_L \tilde{\Phi} Q_R + 
h_\ell \bar{\ell}_L \Phi \ell_R  \nonumber \\
&& + h'_\ell \bar{\ell}_L \tilde\Phi \ell_R+
f{\ell}^{\sf T}_R \Delta_R \ell_R ~+~ {\rm H.c.} \,,
\label{eq:lag}
\end{eqnarray}
where $\tilde{\Phi} = \sigma_2 \Phi^\ast \sigma_2$ (with $\sigma_2$ being the second Pauli matrix), and $h_{q,\,\ell}$, $h_{q,\,\ell}^\prime$, $f$ are the Yukawa coupling matrices. The scalar potential for the scalar sector of the model is\footnote{The most general form of the LRSM scalar potential can be found e.g. in Refs.~\cite{Deshpande:1990ip,Dev:2018xya}.} 
\begin{widetext}
\begin{eqnarray}
V(\Phi, \Delta_R) &=&
-\mu^2_1 {\rm Tr} (\Phi^\dagger\Phi)-\mu^2_2[ {\rm Tr} (\Phi^\dagger\tilde{\Phi} + {\rm Tr} (\Phi\tilde{\Phi}^\dagger)]-\mu^2_3 {\rm Tr} (\Delta^\dagger_R\Delta_R) \nonumber \\ 
&& +\lambda_1 {\rm Tr} (\Phi^\dagger\Phi)^2
+\lambda_2\left[ [ {\rm Tr} (\Phi^\dagger\tilde{\Phi})]^2 +[{\rm Tr} (\Phi\tilde{\Phi}^\dagger)]^2\right] 
+\lambda_3 {\rm Tr} (\Phi^\dagger\tilde{\Phi}) {\rm Tr} (\Phi\tilde{\Phi}^\dagger) 
+\lambda_4 {\rm Tr} (\Phi^\dagger\Phi)
    \left[ {\rm Tr} (\Phi\tilde{\Phi}^\dagger) + {\rm Tr} (\Phi^\dagger\tilde{\Phi})\right] \nonumber \\
&& +\rho_1 \left[ {\rm Tr} (\Delta^\dagger_R \Delta_R)\right]^2 
+ \rho_2 {\rm Tr} (\Delta_R \Delta_R) {\rm Tr} (\Delta^\dagger_R \Delta^\dagger_R) \nonumber \\
&& +\alpha_1 {\rm Tr} (\Phi^\dagger\Phi) {\rm Tr} (\Delta^\dagger_R \Delta_R)+ \left[\alpha_2e^{i\delta_2} \Phi^\dagger\tilde{\Phi } {\rm Tr} (\Delta^\dagger_R \Delta_R)~+~  {\rm H.c.} \right] 
+\alpha_3 {\rm Tr} (\Phi^\dagger\Phi \Delta^\dagger_R\Delta_R) \,,
\end{eqnarray}
\end{widetext}
where all the mass parameters $\mu_{1,2,3}^2$ and the quartic couplings $\lambda_{1,2,3,4}$, $\rho_{1,2}$, $\alpha_{1,2,3}$ are real, and $\delta_2$ the explicit CP-violating phase. The vacuum expectation values (VEVs) of the scalar fields can be written as 
\begin{eqnarray}
\langle \Phi \rangle ~=~\left(\begin{array}{cc} \kappa &0\\0 & \kappa'\end{array}\right) \,, \quad
\langle \Delta_R \rangle ~=~\left(\begin{array}{cc} 0&0\\v_R & 0 \end{array}\right) \,,
\end{eqnarray}
where $\kappa^2 +\kappa^{\prime2} = v_{\rm EW}^2$ with $v_{\rm EW} \simeq 174$ GeV the electroweak VEV, and $v_R$ the right-handed scale. 
The nine scalar quartic couplings $\lambda_{1,2,3,4}$, $\rho_{1,2}$ and $\alpha_{1,2,3}$ together with the VEVs $\kappa, \kappa', v_R$ determine the masses and mixings among the scalars. These are important for the discussions below. The phase $\delta_2$ in the scalar potential is set to zero for simplicity, since it is not relevant for our discussions here.  After symmetry breaking and in the unitary gauge, the remnant physical scalar fields are: the SM Higgs field $h$, the heavy bidoublet fields $H^0_1, A^0_1, H^{\pm}$, the $v_R$-breaking scalar $H^0_3$ which is our candidate field for the 95 GeV excesses, and the doubly-charged scalars $H^{\pm\pm}$. Their masses are given in terms of the couplings above and the field VEVs with the hierarchy $\kappa' \ll \kappa \ll v_R$.\footnote{We expect $\frac{\kappa'}{\kappa}\lesssim \frac{m_b}{m_t}\simeq 0.03$, although strictly, it can be larger if we allow for fine tuning.} To the leading order, the masses of the scalar fields are given by~\cite{Dev:2016dja, Maiezza:2016ybz} 
\begin{align}
m^2_h &\simeq \left( 4\lambda_1 -\frac{\alpha^2_1}{\rho_1}\right)\kappa^2 \,, \\
\label{eqn:mH1}
m^2_{H^0_1} &\simeq m^2_{A^0_1}\simeq m^2_{H^\pm} \simeq \alpha_3v^2_R \,, \\
m^2_{H^0_3} &\simeq 4\rho_1 v^2_R \,, \\
m^2_{H^{\pm\pm}} &\simeq 4 \rho_2 v^2_R \,.
\end{align}
The scalar masses are thus determined by four independent couplings which we adjust to fit observations.  We fix the SM Higgs mass $m_h\simeq 125$ GeV. The new bidoublet scalar masses $m_{H_1^0, A_1^0}$ are constrained to be larger than roughly 15 TeV to satisfy the FCNC constraints from $K-\bar{K}$ and $B-\bar{B}$ transitions, with the right-handed scale $v_R \gtrsim 10$ TeV~\cite{Zhang:2007da,Maiezza:2010ic, Maiezza:2014ala,Bertolini:2019out,Dekens:2021bro}. Similar limits on $v_R$ are obtained from the LHC limits on the $W_R$ mass~\cite{CMS:2021dzb,ATLAS:2023cjo}. It follows from Eq.~(\ref{eqn:mH1}) that the mass of the singly-charged scalar field $ H^\pm$ must be close to the masses of $H^0_1$ and $A_1^0$. Our interest in this paper is in the $SU(2)_R$-breaking scalar $H^0_3$, 
which is identified as the 95 GeV scalar to explain the LHC diphoton and ditau, and the LEP $Zb\bar{b} $ excesses.
 
The interactions and properties of $H^0_3$ have been studied in great details in Refs.~\cite{Dev:2016dja,Dev:2016nfr,Dev:2017dui}. We summarize the relevant results for our purpose below: 

\begin{itemize}

\item  For $v_R\sim 10$ TeV, the quartic coupling $\rho_1$ must be around $2.5\times 10^{-5}$ if $H^0_3$ has to have a mass of ${\cal O}(100\; {\rm GeV})$. This is not such an unnatural value for $\rho_1$, since the one-loop radiative corrections to  this coupling coming from the gauge boson intermediate states is of order ${g^4_R}/{32\pi^2}\sim 10^{-4}$, where $g_R$ is the gauge coupling for $SU(2)_R$ interactions, which is assumed to be equal to $g_L$ for $SU(2)_L$ interactions.

\item The $H^0_3$ scalar mixes with all other CP-even neutral scalars in the model, i.e., the SM Higgs $h$ and the heavy $H^0_1$, with adjustable mixing angles denoted by $\theta_1$ and $\theta_2$, respectively. In terms of the quartic couplings of the model, we have $\sin\theta_1\simeq {\alpha_1\epsilon}/{2\rho_1}$ and $\sin\theta_2 \simeq -{4\alpha_2\epsilon}/{\alpha_3}$ with $\epsilon \equiv {\kappa}/{v_R} \ll 1$. In our fitting below, we will see that the two mixing angles are hierarchical: $\sin\theta_2 \ll \sin\theta_1$.  
This is important, as $\theta_2$ is responsible for the FCNC couplings and should be small, whereas $\theta_1$ needs to be significant in order for $H^0_3$ to fit the 95 GeV excesses.

\item The primary tree-level interactions of $H^0_3$ are with the other scalar fields and the gauge bosons  of the theory. Out of all fermions in the LRSM, it couples only to the right-handed neutrinos but not directly to the charged leptons or quarks. This means that the pure  $H^0_3$ production in $pp$ and $e^+e^-$ collisions arises only via its mixing with the SM Higgs boson $h$. There are other production modes where $H^0_3$ is produced in association with other particles. They are, however, suppressed for the parameter range considered here, and thus not of interest to us in explaining the 95 GeV excesses.

\item $H^0_3$ can directly decay to two photons at the one-loop level via its couplings to $W^\pm_R$ and the charged scalars $H^\pm$ and $H^{\pm\pm}$.  
The induced coupling to photons is inversely proportional to the $v_R$ scale. As a result, for large $v_R$ values, the diphoton decay channel $H_3^0 \to \gamma\gamma$ proceeds predominantly via the $H^0_3-h$ mixing. Thus for a broad range of parameters of interest, both the production and decay of $H^0_3$ proceed via mostly the $H_3^0-h$ mixing. This is true for both $pp$ and $e^+e^-$ collisions. 

\item We also comment here briefly on the FCNC constraints in the LRSM, which are relevant to our fitting below. The primary scalar source of FCNC processes such as the $K-\bar{K}$ and $B-\bar{B}$ mixings are via the tree-level exchange of the $H^0_1$ and $A_1^0$ bosons. We choose the masses of $H^0_1$ and $A_1^0$ to be larger than 15 TeV or so, so that this contribution is at an acceptable level. This requires the quartic coupling $\alpha_3\simeq 3$ in the scalar potential which is still at the perturbative level for $v_R\simeq 10$ TeV. When  $H^0_3$ mass is around 100 GeV, there are also other contributions to FCNCs in the model from $H^0_3$ exchange, due to its mixing with $H^0_1$. This also depends on the VEV ratio $\xi \equiv \kappa'/\kappa$ in the bidoublet sector. 
For $\xi=0$, one can set the $H^0_3-H^0_1$ mixing ${\rm sin}~\theta_2$ equal to zero or very small such that the FCNC effects are under control. On the other hand, for $\xi\neq 0$, there are always FCNC effects from the combination of $\xi {\rm sin}\theta_1+{\rm sin}\theta_2$. Since $\sin\theta_1$ needs to be large for fitting the 95 GeV boson in the model, we can choose $\sin\theta_2$ appropriately to cancel this. More details of the FCNC constraints can be found in Section~\ref{sec:implications}. Unless otherwise specified, we set $\xi=0$ in the following.  

\item Depending on the Yukawa coupling $f$ in Eq.~\eqref{eq:lag}, the right-handed neutrino mass $M_N=fv_R$ could be small, which would bring in additional constraints from lepton number violating (LNV) and/or lepton flavor violating (LFV) processes; see e.g., Refs.~\cite{Tello:2010am,Rodejohann:2011mu,Deppisch:2012vj, Deppisch:2015qwa}. Similarly, relatively light doubly-charged scalars $H^{\pm\pm}$ could give additional contributions to the low-energy LFV and LNV processes~\cite{Bambhaniya:2015ipg}. However, we have enough freedom in the parameter choice to satisfy these constraints, without affecting the light $H_3^0$ phenomenology discussed here. 
\end{itemize}

\section{$H^0_3$ as the 95 GeV boson}
\label{sec:excess}

\begin{table}[t!]
\centering
\caption{Observed signal rates and the local (global) significance for the 95 GeV excess in different channels~\cite{Azevedo:2023zkg}.}
\label{tab:anomalies}
\begin{tabular}{c|c|c}
\hline\hline
Channel & Signal rate $\mu$ & Local (global) sig. \\
\hline
$gg \to H_{95} \to \gamma\gamma$ & $0.33_{-0.12}^{+0.19}$ & $2.9 \ (1.3) \sigma$~\cite{CMS:2023yay}  \\ \hline
$gg \to H_{95} \to \tau^+ \tau^-$ & $1.23_{-0.49}^{+0.61}$ & $2.6 \ (2.3) \sigma$~\cite{CMS:2022goy}  \\ \hline
$e^+ e^- \to Z H_{95} \to Zb\bar{b}$ & $0.117\pm0.057$ & $2.3 \ (<1) \sigma$~\cite{LEPWorkingGroupforHiggsbosonsearches:2003ing} \\ 
\hline\hline
\end{tabular}
\end{table}

The observed signal rates and their local (global) significance from the LHC and LEP data for the 95 GeV  excesses are collected in Table~\ref{tab:anomalies}.  Here the signal rate $\mu$ is defined as the production cross section of the scalar $H_3^0$ times the corresponding branching ratio (BR), with respect to that for a SM-like Higgs boson at 95 GeV, i.e.
\begin{align}
    \mu(H_{95})_X = \frac{\sigma(H_{95})\times {\rm BR}(H_{95}\to X)}{\sigma^{\rm SM}(h_{95})\times {\rm BR}^{\rm SM}(h_{95}\to X)} \, .
    \label{eq:signal}
\end{align}
In this section, we would like to discuss whether the LHC and LEP data corresponding to the 95 GeV boson can be reproduced in the minimal LRSM we are considering here. For this purpose, we need to know the production cross sections of $H^0_3$ in $pp$ and $e^+e^-$ collisions and then the BRs of $H^0_3$ for the $\gamma\gamma$, $b\bar{b}$ and $\tau^+ \tau^-$ modes. There are several production channels for $H^0_3$ in $pp$ collision as discussed in Ref.~\cite{Dev:2016nfr},  with the dominant one being via its mixing with the SM Higgs $h$. In the $e^+e^-$ collision, the mixing with the SM Higgs is the only source. 

The full decay width expression for $H_3^0$ can be found in Refs.~\cite{Dev:2016nfr,Dev:2017dui}. We see from there that, the diphoton channel proceeds both via its mixing with the SM Higgs as well as the one-loop diagrams involving the charged states $W_R^\pm$, $H^\pm$ and $H^{\pm\pm}$ in the LRSM. For low $v_R$ scale (for instance, below 10 TeV) the one-loop contribution is important, which gives the $v_R$ dependence to the signal rates in Eq.~\eqref{eq:signal}. However, the one-loop contribution goes down as $v^2_R$ and becomes subdominant for larger $v_R$, as compared to the SM Higgs mixing contribution. Below, we discuss our detailed numerical fits for three different parameter regions of interest. 

\subsection{The case of $\xi = 0$ and $\sin\theta_2=0$}
\label{sec:xizero}

\begin{figure}[t!]
  \centering
 \includegraphics[width=0.99\linewidth]{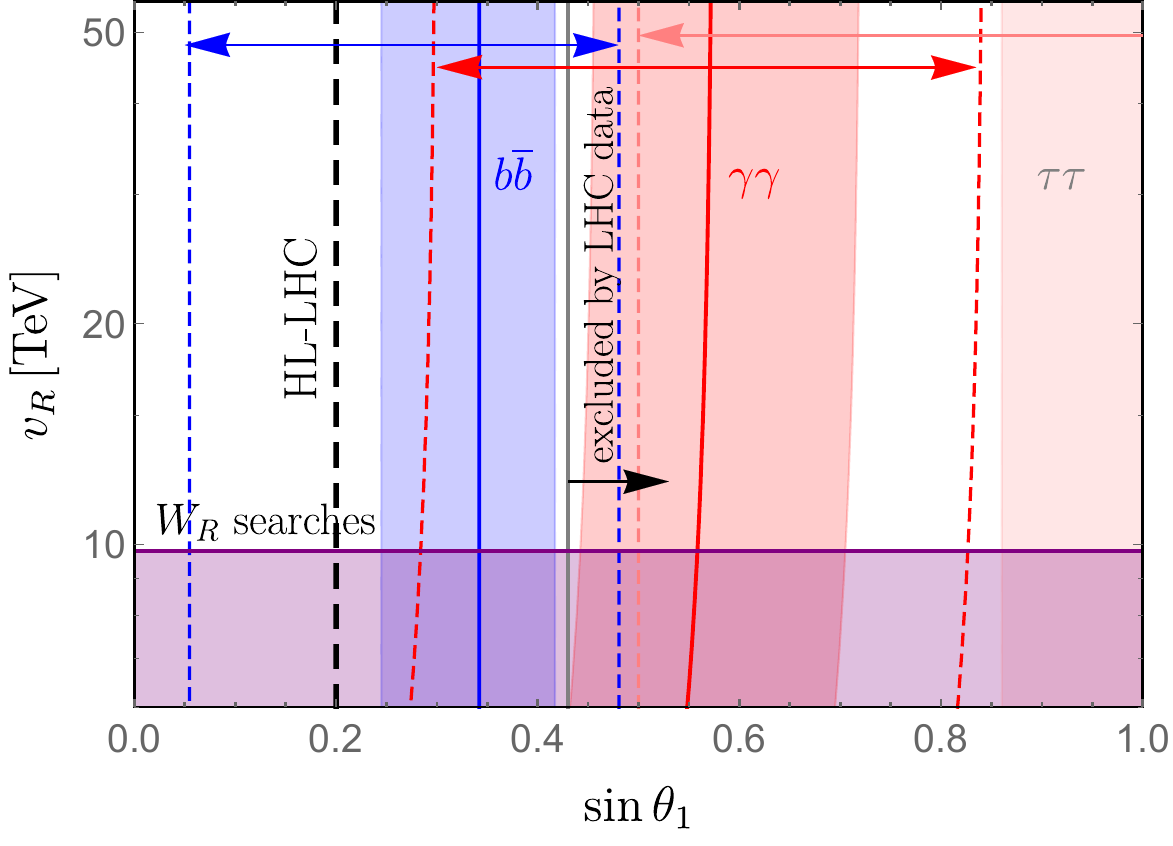} 
  \caption{Preferred range of $\sin\theta_1$ and $v_R$ for the 95 GeV anomalies in the case of $\xi=0$ and $\sin\theta_2=0$ in the $\gamma\gamma$ (red), $b\bar{b}$ (blue) and $\tau^+ \tau^-$ (pink) channels. The central values are indicated by the dashed lines, and the $1\sigma$ and $2\sigma$ ranges are shown, respectively, as the shaded regions and arrows. The current LHC limit on $\sin\theta_1$ and the future prospect at HL-LHC are shown as the vertical solid gray line and dashed black line, respectively~\cite{Falkowski:2015iwa,deBlas:2019rxi}. The horizontal purple shaded region is excluded by the direct searches of $W_R$ in the LHC Run-2~\cite{CMS:2021dzb,ATLAS:2023cjo}. } 
  \label{fig:anomaly:kapp2zero}
  \end{figure}

The fit depends on whether the VEV ratio $\xi\equiv \kappa'/\kappa$ is zero or nonzero. Let us first consider the simpler case of $\xi = 0$. In this case, we can  choose $\sin\theta_2 = 0$, i.e. no $H^0_1-H^0_3$ mixing. Then the couplings of $H^0_3$ with SM fermions are only from its mixing with the SM Higgs, and there is no direct tree-level FCNC couplings via $H_3^0$ exchange. The $\gamma\gamma$, $b\bar{b}$ and $\tau^+\tau^-$ excesses at $1\sigma$ confidence level (C.L.) are shown, respectively, as the blue, red and pink shaded regions in the $\sin\theta_1 - v_R$ plane in Fig.~\ref{fig:anomaly:kapp2zero}. Their central values are indicated by the corresponding solid lines. 

The existing limit on the scalar mixing angle $\sin\theta_1 < 0.44$ at 95\% C.L.~\cite{Falkowski:2015iwa} (see also Refs.~\cite{Robens:2015gla, Robens:2016xkb}) is shown as the vertical gray line. Using the latest LHC measurements of the 125 GeV Higgs boson signal strength, 
\begin{align}
    \mu(h_{125})=\left\{\begin{array}{ll}1.05\pm 0.06 & (\textrm{ATLAS~\cite{ATLAS:2022vkf}})\\
    1.002\pm 0.057 & (\textrm{CMS~\cite{CMS:2022dwd}})
    \end{array}
    \right. ,
\end{align}
which scales as $\cos^2\theta_1$ in our model, we derive a more stringent bound on 
\begin{align}
    \sin\theta_1 \lesssim \left\{\begin{array}{ll}
0.26 & (\textrm{ATLAS})\\
0.33 & (\textrm{CMS})
    \end{array}\right. \quad \textrm{at 95\% C.L.}
    \label{eq:run2}
\end{align}
However, this assumes that all production and decay processes of the 125 GeV Higgs boson scale with the same global signal strength. A dedicated analysis of the Higgs data, taking into account the experimental measurements of the individual production and decay channel signal strengths, is beyond the scope of the current work. Therefore, we will simply use the old bound of 0.44~\cite{Falkowski:2015iwa} in the following analysis. The future limit on  $\sin\theta_1$ at the high-luminosity LHC (HL-LHC) can reach down to 0.2~\cite{deBlas:2019rxi}, as indicated by the vertical dashed black line. 

One should note that, although the FCNC couplings are absent in the limit of $\xi = 0$ and $\sin\theta_2 = 0$, the one-loop couplings of $H_3^0$ with photons through the heavy charged scalar and gauge bosons are still there, and diphoton decay width $\Gamma (H_3^0 \to \gamma\gamma)$  depends on the $v_R$ scale, as can be seen from the non-vertical nature of the red curves in  Fig.~\ref{fig:anomaly:kapp2zero}. The current most stringent limit on the $v_R$ scale is from the direct searches of $W_R$ at the LHC Run-2, which have excluded the $W_R$ mass below 6.4 TeV~\cite{ATLAS:2023cjo}. Assuming $g_R = g_L$, this corresponds to a constraint on $v_R > 9.8$ TeV, which is indicated by the horizontal purple shaded region in Fig.~\ref{fig:anomaly:kapp2zero}. In principle, the signal rates of $b\bar{b}$ and $\tau^+ \tau^-$ channels should also depend on the $v_R$ scale through the corresponding BRs, which depend on the total width of $H_3^0$ (and hence, on the $\gamma\gamma$ partial width). However, for $H_3^0$ at 95 GeV with a large $\sin\theta_1$ as required for the anomalies, the ${\rm BR} (H_3^0 \to \gamma\gamma)$ is so small (see Fig.~\ref{fig:BR:kappa2zero} below) that the dependence on the $v_R$ scale in the diphoton channel hardly affects the other decay channels of $H_3^0$. Therefore, the preferred regions for $b\bar{b}$ and $\tau^+\tau^-$ anomalies are almost vertical in Fig.~\ref{fig:anomaly:kapp2zero}. 

  \begin{figure*}[t!]
\centering
\includegraphics[width=0.49\linewidth]{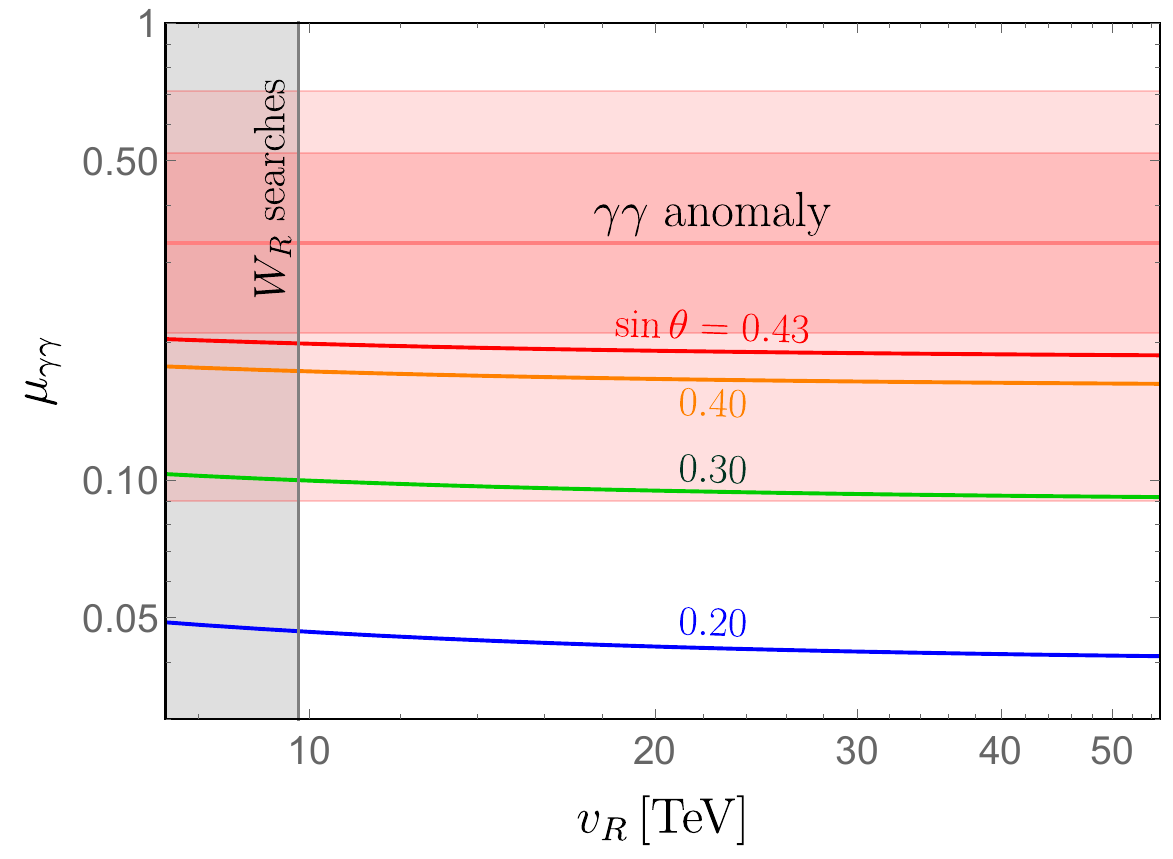} 
\includegraphics[width=0.49\linewidth]{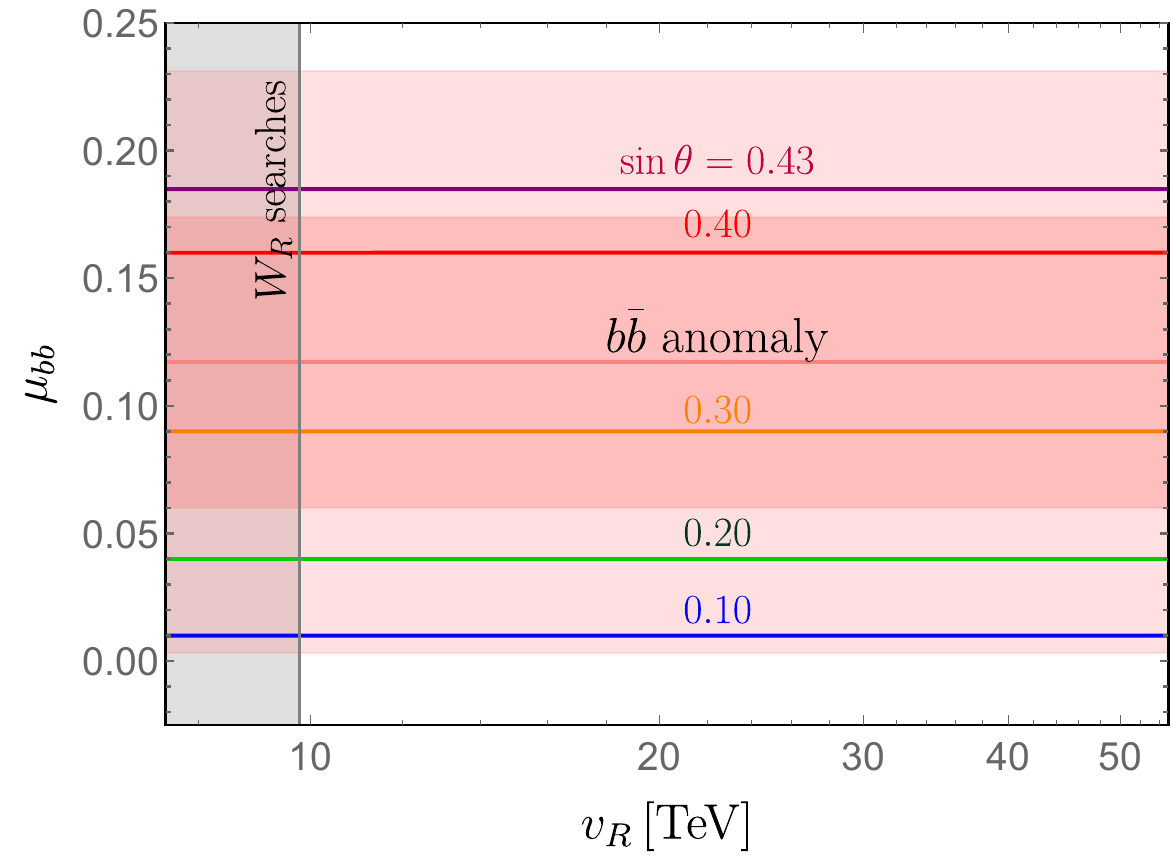} 
\caption{Signal rates of the 95 GeV anomalies in the $\gamma\gamma$ (left) and $b\bar{b}$ (right) channels for the case of $\xi=0$ and $\sin\theta_2=0$ as functions of the $v_R$ scale, for different values of the $h-H_3^0$ mixing angle $\sin\theta_1$. 
The darker (lighter) pink bands are the corresponding $1\sigma$ ($2\sigma$) C.L. preferred ranges. The vertical gray shaded region is the LHC exclusion from $W_R$ searches.
} 
\label{fig:anomaly:vR:kapp2zero}
\end{figure*}

We see from Fig.~\ref{fig:anomaly:kapp2zero} that for the simplest case of $\xi = 0$ and $\sin\theta_2=0$, the preferred regions for the $\gamma\gamma$, $b\bar{b}$ and $\tau^+ \tau^-$ anomalies do not overlap at the $1\sigma$ C.L. This can be understood as follows:
\begin{itemize}
    \item In the limit of large $v_R$ (decoupling all heavy particles in the LRSM except $H_3^0$), and $\xi = 0=\sin\theta_2$ case, the $H_3^0$ scalar is much like the extra scalar in the singlet extension of SM in the scalar sector, i.e., all the couplings of $H_3^0$ to the SM particles are rescaled by its mixing with the SM Higgs boson.
    \item For a SM-like Higgs boson at 95 GeV, the diphoton BR is only 0.0014~\cite{LHCHiggsCrossSectionWorkingGroup:2016ypw}, but the diphoton anomaly rate is $0.33_{-0.12}^{+0.19}$ [cf.~Table~\ref{tab:anomalies}], with respect to that for the SM-like Higgs. This means that the mixing angle is required to be large: $\sin\theta_{\gamma\gamma}\in [0.46,0.67]$ at $1\sigma$ (red shaded region in Fig.~\ref{fig:anomaly:kapp2zero}). 
    
    \item The $b\bar{b}$ BR in the SM-like Higgs case is large, roughly 0.82 at 95 GeV~\cite{LHCHiggsCrossSectionWorkingGroup:2016ypw}. However, the $b\bar{b}$ anomaly rate is only $0.117 \pm 0.057$ [cf.~Table~\ref{tab:anomalies}], which implies that the required mixing angle is relatively small: $\sin\theta_{b\bar{b}}\in [0.24,0.42]$ at $1\sigma$ (blue shaded region in Fig.~\ref{fig:anomaly:kapp2zero}).

    \item For a SM-like Higgs at 95 GeV, the BR for the $\tau^+ \tau^-$ channel is only 0.08~\cite{LHCHiggsCrossSectionWorkingGroup:2016ypw}, but the observed signal rate is large in this channel, with central value larger than one [cf.~Table~\ref{tab:anomalies}]. With only a singlet scalar extension of the SM, it is therefore impossible to accommodate the central value without any other enhancement in the $\tau^+\tau^-$ sector. However, the uncertainty in this channel is pretty large, and the $1\sigma$ C.L. region $\sin\theta_{\tau^+\tau^-}\in [0.86,1]$ (pink shaded region in Fig.~\ref{fig:anomaly:kapp2zero}).
\end{itemize}
Therefore, with $\xi=0=\sin\theta_2$ which is effectively like the singlet scalar extension of the SM, there is no overlap between the preferred regions for $\gamma\gamma$, $b\bar{b}$ and $\tau^+ \tau^-$ anomalies at $1\sigma$ C.L.. 
At the $2\sigma$ C.L., however, a broader range of $\sin\theta_1$ values can be accommodated. To this end, the $2\sigma$ bands for the $\gamma\gamma$, $b\bar{b}$ and $\tau^+ \tau^-$ anomalies are indicated by the arrows in Fig.~\ref{fig:anomaly:kapp2zero}. It is clear that at $2\sigma$ C.L. there is sizable overlap for the $\gamma\gamma$ and $b\bar{b}$ bands, as well as the $\gamma\gamma$ and $\tau^+\tau^-$ bands (which is however excluded by LHC Higgs data). The entire $2\sigma$ region for the $\gamma\gamma$ excess is on the verge of exclusion from the Run-2 LHC data if we take the na\"{i}ve limit from Eq.~\eqref{eq:run2}, and is completely within reach of HL-LHC. In any case, we can not simultaneously accommodate all three anomalies even at the $2\sigma$ C.L.

In light of the unusually large signal rate (and uncertainties) in the $\tau^+ \tau^-$ channel and the difficulties in fitting it, we will focus more on the $\gamma\gamma$ and $b\bar{b}$ channels in the following. More details of the $\gamma\gamma$ and $b\bar{b}$ anomalies in the case of $\xi = 0$ and $\sin\theta_2=0$ are presented in Fig.~\ref{fig:anomaly:vR:kapp2zero}. The left and right panels are respectively for the signal rates $\mu_{\gamma\gamma}$ and $\mu_{b\bar{b}}$, as functions of the $v_R$ scale. In both panels, the horizontal pink lines indicate the central values for the signal rates, and the darker (lighter) shaded regions are the $1\sigma$ ($2\sigma$) C.L. ranges. In the left panel, the red, orange, green and blue lines denote, respectively, the  values of $\sin\theta_1 = 0.43$ (the LHC upper limit), 0.40, 0.30 and 0.20 for the signal rate as function of the $v_R$ scale, while in the right panel, the purple, red, orange, green and blue lines are respectively for the values of $\sin\theta_1 = 0.43$, 0.40, 0.30, 0.20 and 0.10. It is clear from Fig.~\ref{fig:anomaly:vR:kapp2zero} that the mixing $0.30 \lesssim \sin\theta_1 < 0.44$ can  accommodate the diphoton signal at $2\sigma$ C.L., with weak dependence on the $v_R$ scale. while for the $b\bar{b}$ excess the mixing is required to be within the range of $0.10 \lesssim \sin\theta_1 <0.44$.

\subsection{The case of $\xi = 0$ and $\sin\theta_2 \neq 0$}
\label{sec:optimal}

In the case of $\xi \neq 0$ and/or $\sin\theta_2 \neq 0$, the couplings of $H^0_3$ with SM quarks are from a combination of those from its mixing with the SM Higgs through $\sin\theta_1$ and its mixing with the heavy scalar $H_1^0$ through $\sin\theta_2$. For instance, the couplings of $H_3^0$ to the up and down-type quarks are given by~\cite{Dev:2016dja,Dev:2016nfr,Dev:2017dui}
\begin{eqnarray}
\label{eqn:Yukawacoupling}
H_3^0u\bar{u}: & \quad & \hat{Y}_U \sin\tilde\theta_1 - \Big( V_L \hat{Y}_D V_R^\dagger \Big) \sin\tilde\theta_2 \,, \\
H_3^0d\bar{d}: & \quad & \hat{Y}_D \sin\tilde\theta_1 - \Big( V_L^\dag \hat{Y}_U V_R \Big) \sin\tilde\theta_2 \,,
\label{eqn:Yukawacoupling1}
\end{eqnarray}
with $\hat{Y}_{U,\,D}$ the diagonal Yukawa coupling matrices in the up- and down-type quark sectors, $V_{L,\,R}$ the left- and right-handed quark mixing matrices, and
\begin{eqnarray}
    \sin\tilde\theta_{1,\,2} = \sin\theta_{1,\,2} + \xi \sin\theta_{2,\,1} \,.
\end{eqnarray}
The second term in Eqs.~(\ref{eqn:Yukawacoupling}) and (\ref{eqn:Yukawacoupling1}) originates from the tree-level FCNC couplings of $H_1^0$ in the LRSM, and the VEV ratio $\xi$ dictates the mixing between the SM Higgs $h$ and the heavy $H_1^0$. This leads to tree-level FCNC processes such as the $K - \bar{K}$ and $B_{d,\,s}-\bar{B}_{d,\,s}$ mixing via exchange of $H^0_3$. Unlike the case in Section~\ref{sec:xizero}, $H_3^0$ might also have new significant flavor-changing decay channels such as $H_3^0 \to b\bar{d} + d\bar{b},\, b\bar{s} + s\bar{b}$. More details on the implications for $H_3^0$ at 95 GeV are given in Section~\ref{sec:implications}.

 \begin{figure}[t!]
  \centering
 \includegraphics[width=0.99\linewidth]{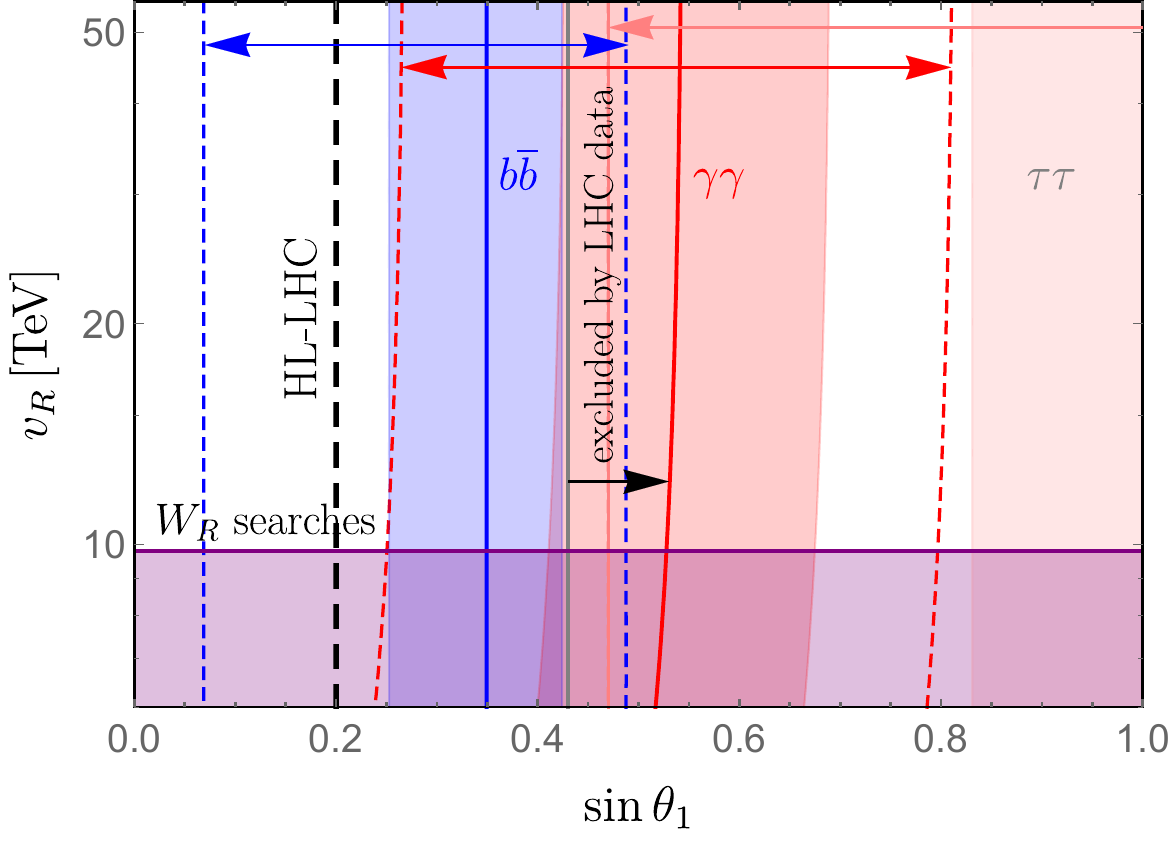} 
  \caption{Same as in  Fig.~\ref{fig:anomaly:kapp2zero}, but for the optimal case of $\xi = 0$ and $\sin\theta_2 = 8.54 \times 10^{-4}$. See text for more details.} 
  \label{fig:anomaly:kapp2nonzero}
  \end{figure}

  \begin{figure*}[!htb]
\centering
\includegraphics[width=0.49\linewidth]{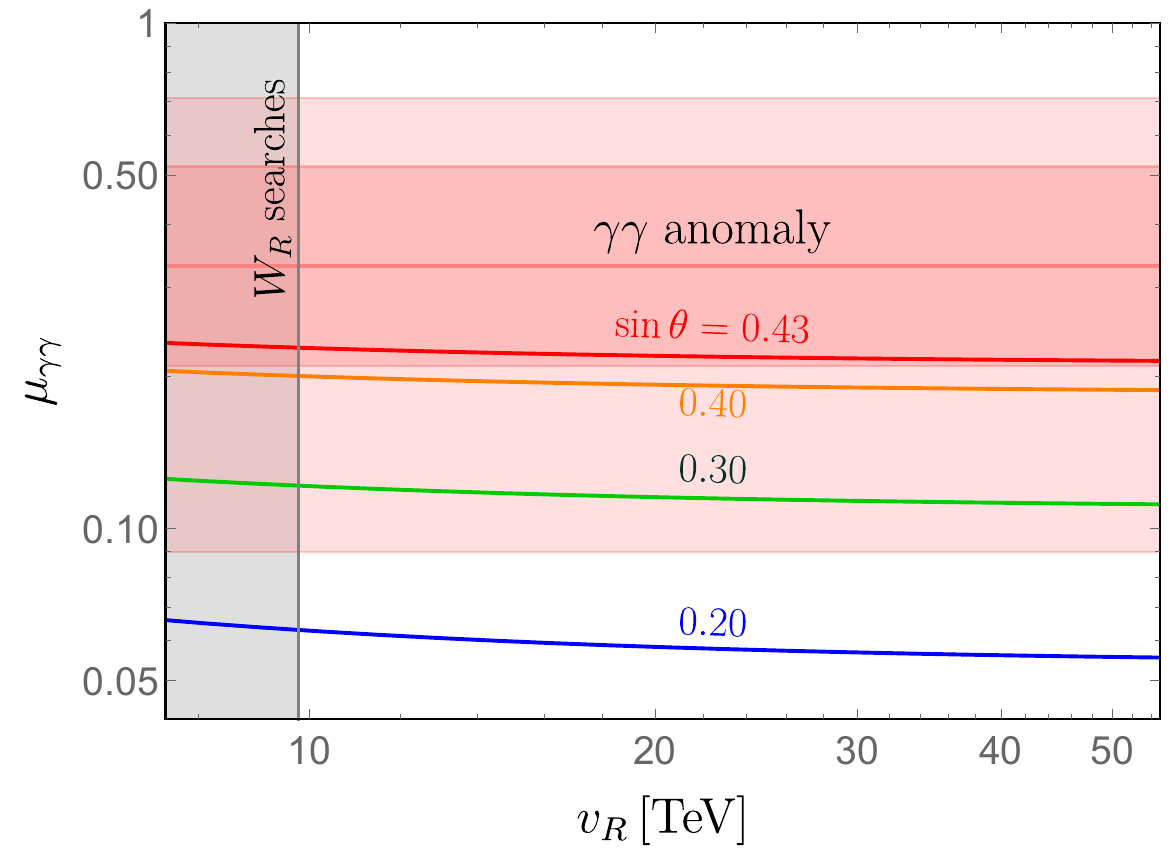} 
\includegraphics[width=0.49\linewidth]{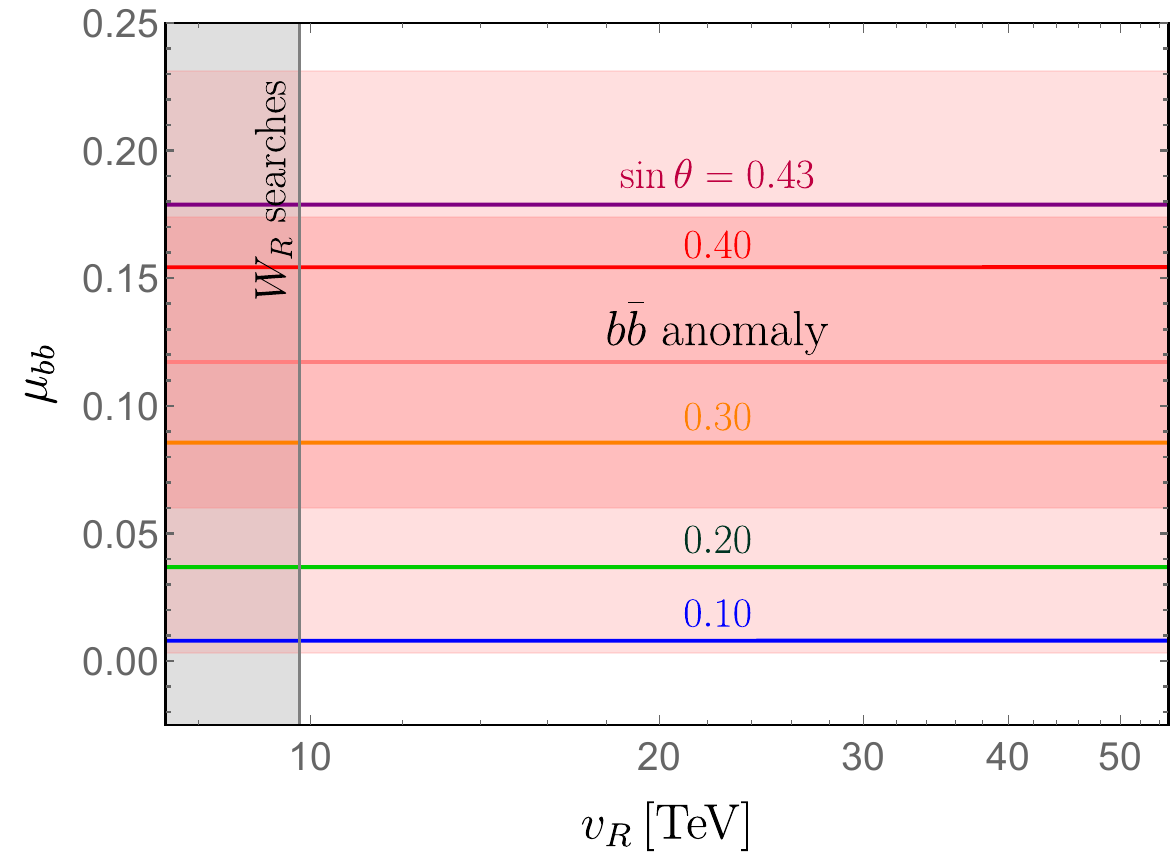} 
\caption{Same as in Fig.~\ref{fig:anomaly:vR:kapp2zero}, but for the optimal case of $\xi = 0$ and $\sin\theta_2 = 8.54 \times 10^{-4}$. See text for more details.} 
\label{fig:anomaly:vR:kapp2nonzero}
\end{figure*}
\begin{figure*}[!htb]
\centering
\includegraphics[width=0.49\linewidth]{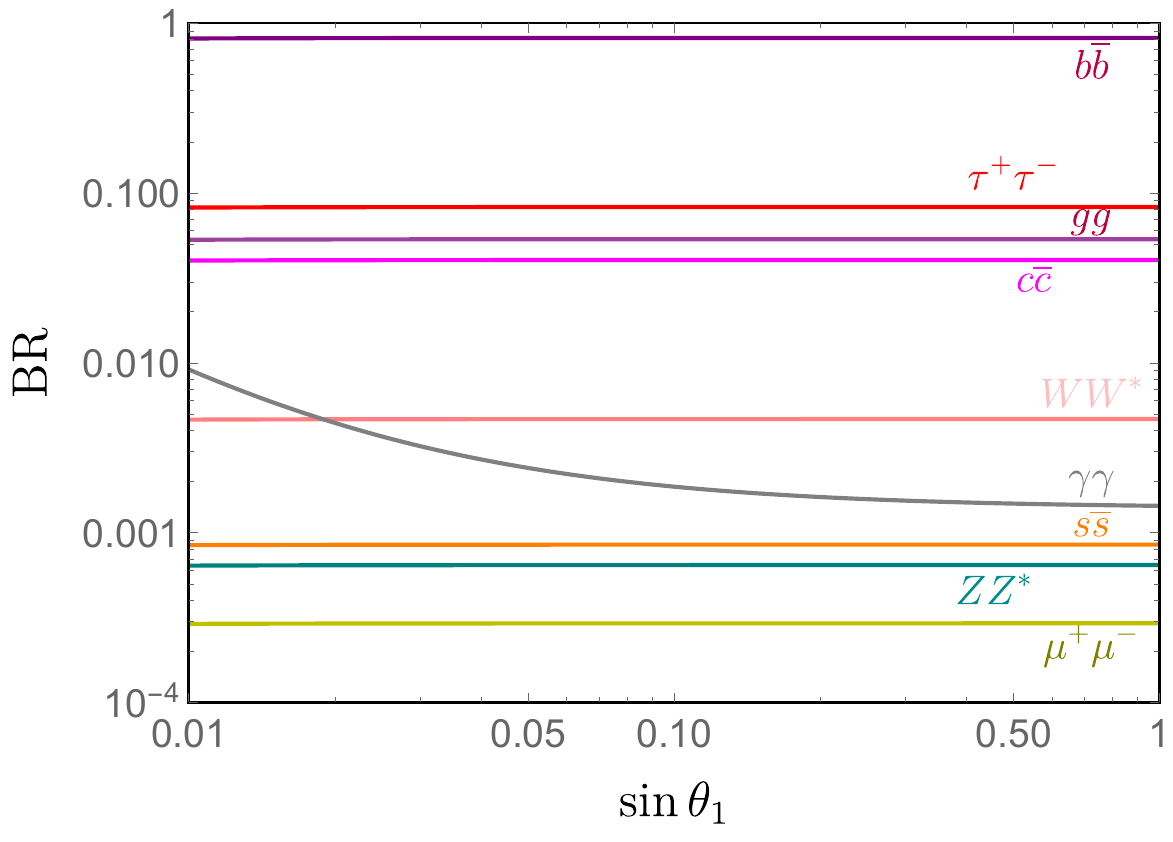} 
\includegraphics[width=0.49\linewidth]{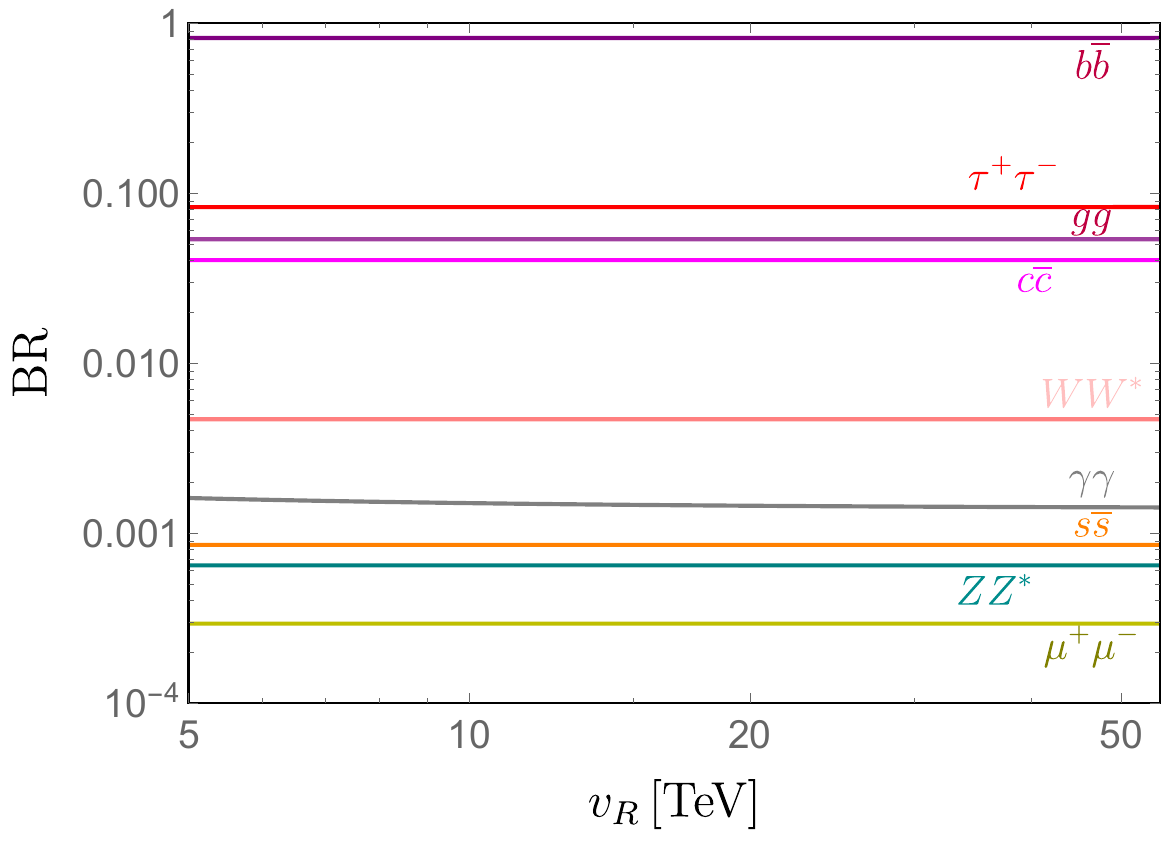} \\
\includegraphics[width=0.49\linewidth]{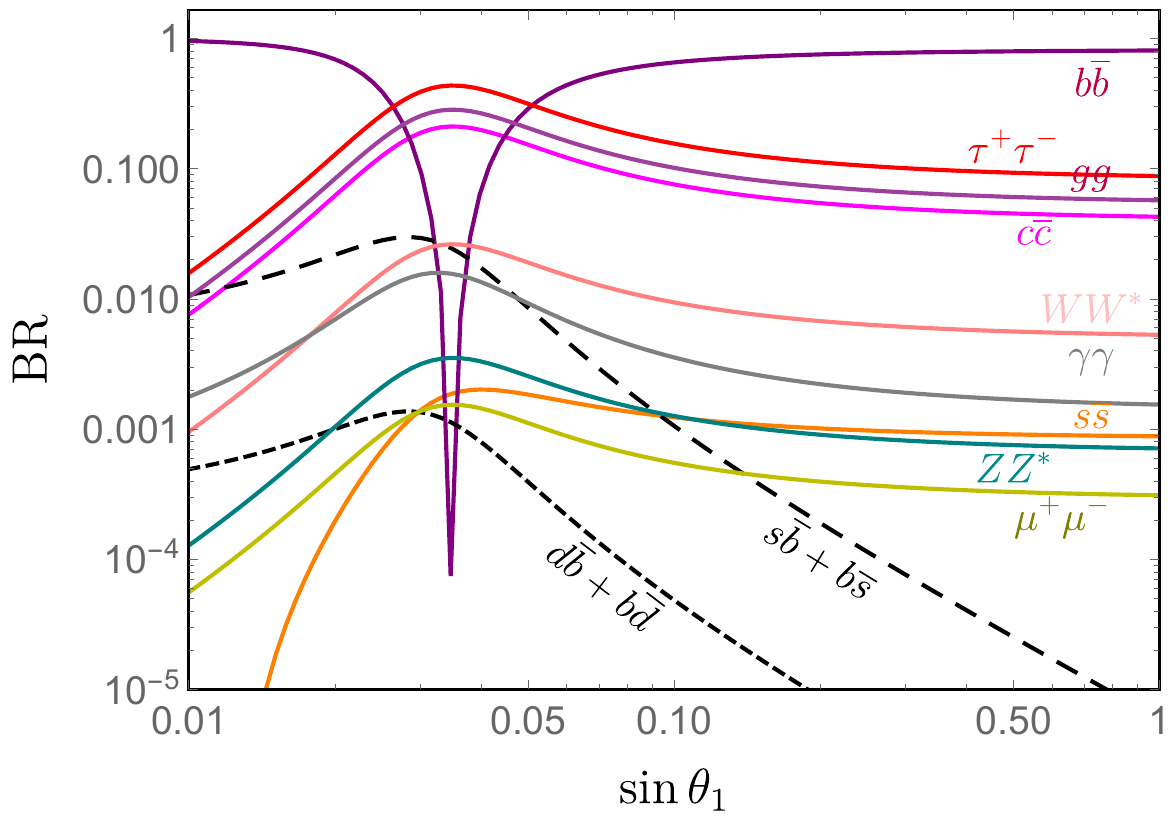} 
\includegraphics[width=0.49\linewidth]{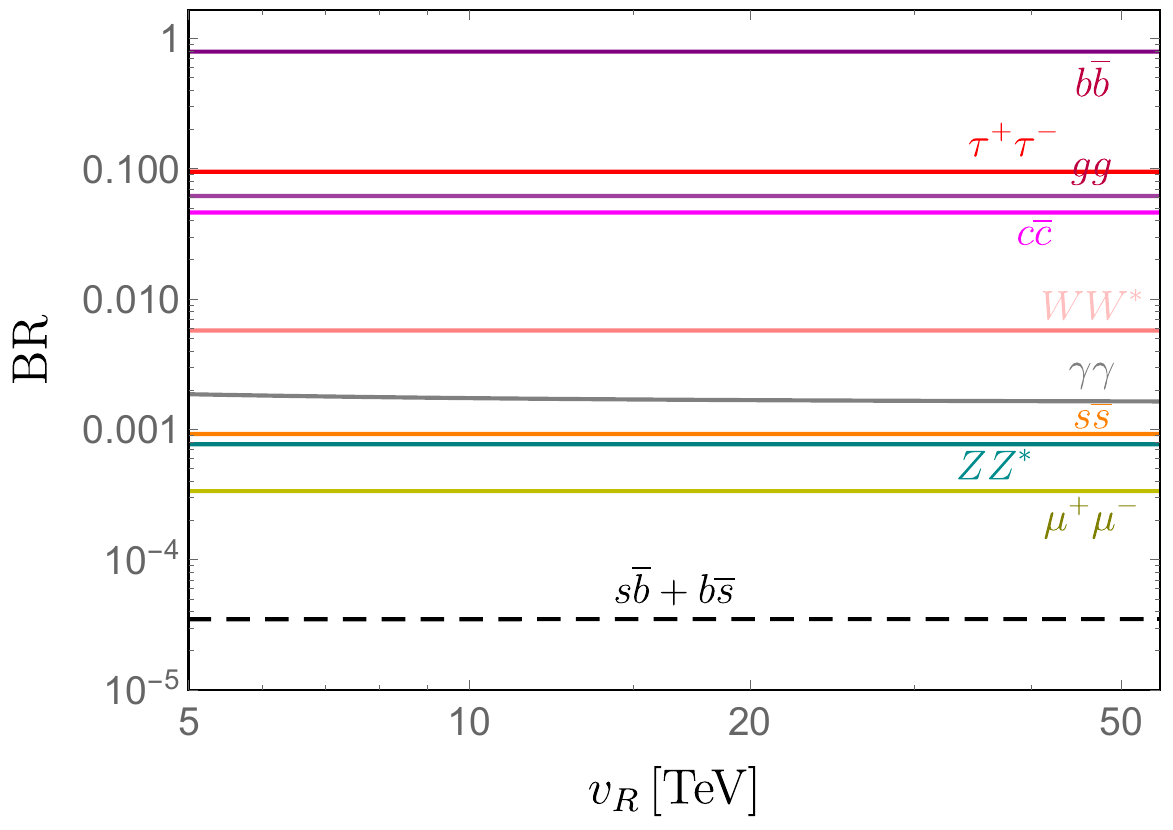} 
\caption{BRs of $H_3^0$ decay as functions of $\sin\theta_1$ with $v_R = 10$ TeV (left) and as functions of $v_R$ with $\sin\theta_1 = 0.4$ (right). The top panels are for the case of $\xi=0$ and $\sin\theta_2=0$, while the bottom panels are for the optimal case of $\xi = 0$ and $\sin\theta_2 = 8.54 \times 10^{-4}$.} 
\label{fig:BR:kappa2zero}
\end{figure*}

For illustration purpose, here we choose a simple benchmark case of $\xi = 0$ but with a non-zero $\sin\theta_2 = 8.54 \times 10^{-4}$. In this case, Eqs.~(\ref{eqn:Yukawacoupling}) and (\ref{eqn:Yukawacoupling1}) are simplified as follows:  
\begin{eqnarray}
\label{eqn:Yukawacoupling2}
H_3^0u\bar{u}: \quad \hat{Y}_U \sin\theta_1 - \Big( V_L \hat{Y}_D V_R^\dagger \Big) \sin\theta_2 \,, \\
H_3^0d\bar{d}: \quad \hat{Y}_D \sin\theta_1 - \Big( V_L^\dag \hat{Y}_U V_R \Big) \sin\theta_2 \,,
\label{eqn:Yukawacoupling3}
\end{eqnarray}
and the FCNC couplings of $H_3^0$ are completely from the $H_3^0 - H_1^0$ mixing. With both contributions to the Yukawa couplings of $H_3^0$ in Eq.~(\ref{eqn:Yukawacoupling3}), we can obtain better fitting of the $\gamma\gamma$ and $b\bar{b}$ anomalies than in the case of $\sin\theta_2 = 0$ in Section~\ref{sec:xizero}. This is shown in Fig.~\ref{fig:anomaly:kapp2nonzero}. The notations are the same as in Fig.~\ref{fig:anomaly:kapp2zero}. The value of $\sin\theta_2$ is chosen to have the optimal overlapping of the $\gamma\gamma$ and $b\bar{b}$ anomalies at $v_R = 10$ TeV while satisfying the FCNC constraints and the LHC limits on $W_R$ mass. For the optimal case here, the $\gamma\gamma$ and $b\bar{b}$ anomalies can be fitted simultaneously at the $1\sigma$ C.L. for $v_R = 10$ TeV, with the mixing angle $\sin\theta_1 \simeq 0.42$. Even the $\tau^+\tau^-$ excess can be fitted simultaneously with the $\gamma\gamma$ and $b\bar{b}$ excesses at 2$\sigma$ C.L.; however, this requires a slightly larger $\sin\theta_1\simeq 0.48$ that is excluded by the LHC Higgs data~\cite{Falkowski:2015iwa}. 

More details of the signal rates $\mu_{\gamma\gamma}$ and $\mu_{b\bar{b}}$ as functions of $v_R$ and $\sin\theta_1$ are presented in Fig.~\ref{fig:anomaly:vR:kapp2nonzero}. In this figure, it is obvious that a mixing angle of $0.26 \lesssim \sin\theta_1 <0.43$ can fit the diphoton excess at the $2\sigma$ C.L. with weak dependence on the $v_R$ scale, while the $b\bar{b}$ anomaly needs the mixing angle to be within $0.10 \lesssim \sin\theta_1 <0.43$, with almost no dependence on the $v_R$ scale.

\subsection{The case of $\xi \neq 0$ and $\sin\theta_2 \neq 0$}

If we set both $\xi$ and $\sin\theta_2$ to be nonzero, one might expect a better fit of the $\gamma\gamma$ and $b\bar{b}$ anomalies. However, in light of the stringent FCNC constraints on the couplings of $H_3^0$, we find that the general case of $\xi \neq 0$ and $\sin\theta_2 \neq 0$ does not provide a better fit of the anomalies than the case of $\xi = 0$ and $\sin\theta_2 \neq 0$. Therefore, we will not present the results for this case. More details on the flavor-changing decays of $H_3^0$ and the FCNC constraints are given in the following section.
  
\section{Implications of identifying $H^0_3$ with 95 GeV Anomalies} 
\label{sec:implications}

In the limit of $v_R \to \infty$, $\xi = 0$ and $\sin\theta_2 = 0$, the scalar $H_3^0$ behaves much like a pure singlet scalar extension of the SM. However, there are some non-trivial features of $H_3^0$ in the LRSM when we consider more general parameter space with either smaller $v_R$  or $\xi\neq 0$ or $\sin\theta_2\neq 0$. In fact, this makes our $H_3^0$ very different from the `trivial' singlet scalar case, and has some phenomenologically interesting implications for the properties of $H_3^0$. For instance, as aforementioned in previous sections, the decay $H_3^0 \to \gamma\gamma$ receives contributions from both its couplings with SM fermions and $W$ boson via mixing with the SM Higgs, as well as from its couplings to the heavy charged bosons in the LRSM. As a result, the partial decay width $\Gamma (H_3^0 \to \gamma\gamma)$, and hence, all the decay BRs, depend on the $v_R$ scale.

For illustration purpose, the decay BRs of $H_3^0$ at 95 GeV in the case of $\xi = 0$ and $\sin\theta_2 = 0$ are shown in Fig.~\ref{fig:BR:kappa2zero} top panels.  The dependence of the BRs on the mixing angle $\sin\theta_1$ is presented in the left panel, where we have set the scale $v_R = 10$ TeV, while the dependence of BRs on the $v_R$ scale is shown in the right panel with a fixed $\sin\theta_1 = 0.4$. In both panels, we have shown only the channels with BR larger than $10^{-4}$, including $H_3^0 \to b\bar{b}$, $\tau^+ \tau^-$, $gg$, $c\bar{c}$, $WW^\ast$, $\gamma\gamma$, $s\bar{s}$, $ZZ^\ast$, $\mu^+ \mu^-$. As can be clearly seen in the left panel of Fig.~\ref{fig:BR:kappa2zero}, the contribution of heavy bosons in the LRSM to the decay $H_3^0 \to \gamma\gamma$ is important only when the mixing angle $\sin\theta_1 \lesssim 0.1$. For $\sin\theta_1 \gtrsim 0.01$, the $v_R$ dependence of all other channels except the diphoton channel can be safely neglected.

The decay BRs of $H_3^0$ in the optimal case of $\xi = 0$ and $\sin\theta_2 = 8.54\times10^{-4}$ are depicted in Fig.~\ref{fig:BR:kappa2zero} bottom panels. As a result of the nonzero $\sin\theta_2$, $H_3^0$ mixes with $H_1^0$ and will have tree-level FCNC couplings to the SM quarks. Furthermore, for the flavor-conserving couplings $H_3^0 u_i \bar{u}_i$ and $H_3^0 d_i \bar{d}_i$ (with $i = 1,\,2,\,3$ the generation index), the two contributions in Eq.~(\ref{eqn:Yukawacoupling3}) might (partially) cancel with each other, making the corresponding Yukawa couplings significantly suppressed. 
This can be seen from the bottom left panel in Fig.~\ref{fig:BR:kappa2zero}, where the vanishing of the $b\bar{b}$ channel at $\sin\theta_1 \simeq 0.035$ is due to this cancellation effect. We can also see that, for fixed $v_R$, the BRs for the FCNC channels $H_3^0 \to s\bar{b} + b\bar{s},\,  d\bar{b} + b\bar{d}$ are very sensitive to the mixing angle $\sin\theta_1$. The BR for $s\bar{b} + b\bar{s}$ can even go up to ${\cal O} (0.01)$. For fixed $\sin\theta_1 = 0.4$ (bottom right panel), the BR for the $s\bar{b} + b\bar{s}$ channel is roughly $3\times10^{-5}$. The BR for the $d\bar{b} + b\bar{d}$ channel is smaller than $10^{-5}$, and is thus not shown here.

\begin{figure}[!t]
\centering
\includegraphics[width=0.99\linewidth]{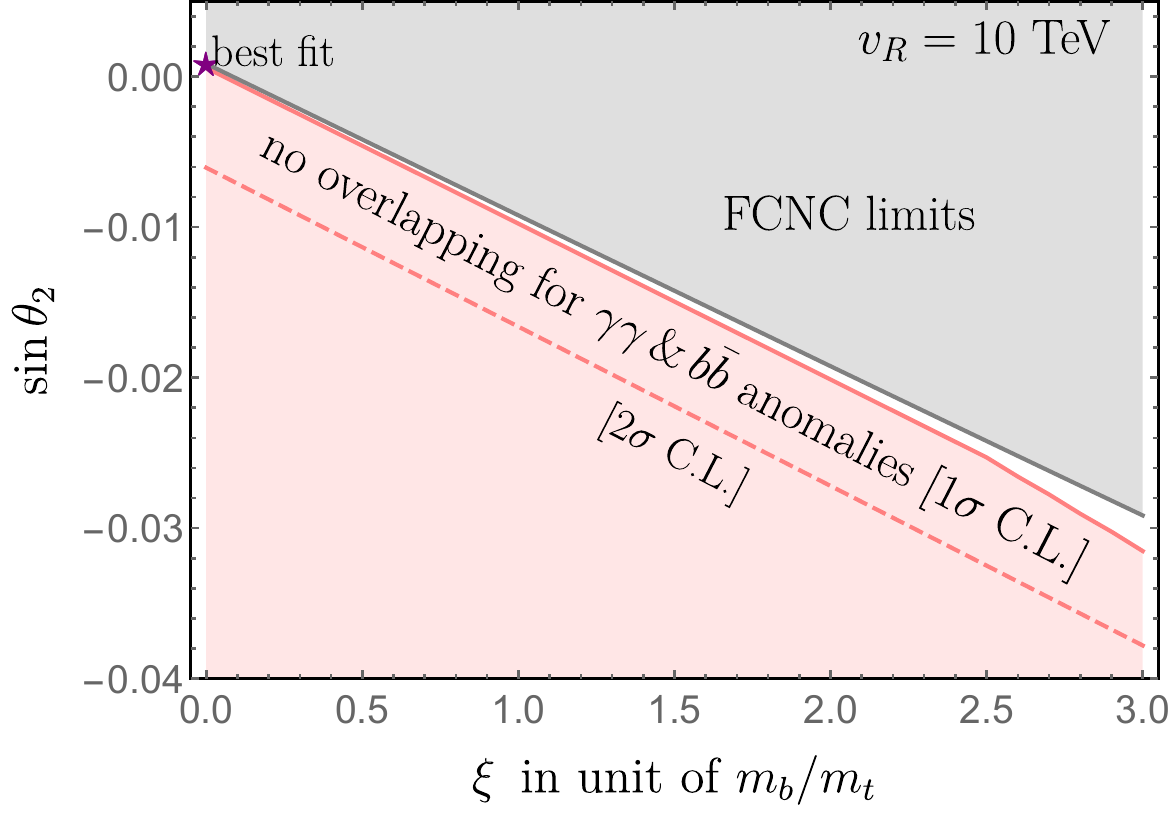} 
\caption{Fitting the 95 GeV $\gamma\gamma$ and $b\bar{b}$ excesses in the LRSM in the parameter space of $\xi$ in unit of $m_b/m_t$ and $\sin\theta_2$. The pink shaded region with solid (dashed) boundary indicates the $1\sigma$ ($2\sigma$) C.L. region where both anomalies cannot be fitted simultaneously. The gray shaded region is excluded by the FCNC constraints on $H_3^0$. 
The best-fit point ($\star$) corresponds to the optimal case of $\xi=0$ and $\sin\theta_2 = 8.54\times10^{-4}$ in Section~\ref{sec:optimal}. We have set $v_R = 10$ TeV. } 
\label{fig:limits}
\end{figure}

As a result of its mixing with $H_1^0$, the couplings of $H_3^0$ in the LRSM are subject to the FCNC constraints. Fixing $m_{H_3^0} = 95$ GeV, the $K - \bar{K}$ and $B - \bar{B}$ mixing data can be used to set limits on the VEV ratio $\xi = \kappa'/\kappa$ and the mixing angles $\sin\theta_{1,\,2}$. Following the calculations in Ref.~\cite{Zhang:2007da}, the FCNC constraints on $\xi$ and $\sin\theta_2$ are depicted as the gray shaded region in Fig.~\ref{fig:limits}, where we have scanned the whole parameter space of $\xi$ and $\sin\theta_{1,\,2}$. It is understandable that when the VEV ratio $\xi$ becomes large, one needs to make the absolute value of $\sin\theta_2$ larger, such that the two terms in $\xi  \sin\theta_1 + \sin\theta_2$ for FCNC couplings cancel out with each other. 
In the pink shaded region in Fig.~\ref{fig:limits}, there is no overlap between the $\gamma\gamma$ and $b\bar{b}$ anomalies at the $1\sigma$ C.L., with $v_R = 10$ TeV. This implies that, to fit both the $\gamma\gamma$ and $b\bar{b}$ anomalies within the $1\sigma$ C.L. without violating the FCNC constraints, the values of $\xi$ and $\sin\theta_2$ have to be within the narrow uncolored band in Fig.~\ref{fig:limits}, with the best-fit point indicated by the star, which corresponds to the optimal case discussed in Section~\ref{sec:optimal}. As exemplified in Figs.~\ref{fig:anomaly:kapp2zero} to \ref{fig:anomaly:vR:kapp2nonzero}, the parameter space for the overlap of diphoton and $b\bar{b}$ excesses at the $2\sigma$ C.L. are much broader. Correspondingly, the allowed boundary for fitting the diphoton and $b\bar{b}$ anomalies at the $2\sigma$ C.L. in the $\xi - \sin\theta_2$ plane is indicated by the pink dashed line.

Since $H^0_3$ in our model is lighter than the SM Higgs, the SM Higgs $h$ can have a rare three-body decay mode $h\to H^0_3h^*\to H^0_3 b\bar{b}$, mediated by the $H_3^0hh$ coupling. The calculational details for the partial width $\Gamma (h \to H_3^0 b\bar{b})$ are given in Appendix~\ref{sec:appendix}. It turns out that for $m_{H_3^0}=95$ GeV, this channel is suppressed by the phase space factor and the off-shell mediator $h$, and has a BR of only ${\cal O} (10^{-8})$. It is therefore unlikely to be observable even at the future precision Higgs factories~\cite{ILC:2013jhg, Abramowicz:2016zbo, An:2018dwb, FCC:2018evy, AlAli:2021let}.  

From Fig.~\ref{fig:anomaly:kapp2nonzero}, we see that the $h-H^0_3$ mixing angle ${\rm sin}\theta_1$ has to be large, of order of $0.4$ or so, for the minimal LRSM to simultaneously explain the 95 GeV excesses in $\gamma\gamma$ and $b\bar{b}$ channels at $1\sigma$ C.L. This mixing value is marginally compatible with the current LHC Higgs constraint~\cite{Falkowski:2015iwa}, and  can be probed both at HL-LHC~\cite{deBlas:2019rxi}, as well as future precision Higgs factories such as ILC~\cite{ILC:2013jhg}, CLIC~\cite{Abramowicz:2016zbo}, CEPC~\cite{An:2018dwb}, FCC-ee~\cite{FCC:2018evy} and muon collider~\cite{AlAli:2021let}. This provides a concrete way to test the LRSM interpretation of the anomalies proposed here, should they persist with more data. Moreover, since the diphoton decay channel of $H_3^0$ depends on the $v_R$ scale, if the 95 GeV excess in the diphoton channel is confirmed with high significance, then we can in principle determine the $v_R$ scale in the minimal LRSM, or at least provide some information on it. This would be complementary to the direct $W_R$ boson  searches at the LHC.
Furthermore, as illustrated in Fig.~\ref{fig:BR:kappa2zero} bottom panels, the BRs for the flavor-changing decays $H_3^0 \to s\bar{b} + b\bar{s},\,  d\bar{b} + b\bar{d}$ could be sizable in some regions of parameter space in the LRSM. The direct searches of these channels at future colliders would provide further independent tests of our LRSM interpretation of the 95 GeV excesses.

\section {Conclusions} 
\label{sec:conclusion}
In this paper, we have pointed out that the minimal LRSM for neutrino masses contains a natural candidate that can play the role of a new Higgs boson at 95 GeV to explain the $\gamma\gamma$ excesses reported in both CMS and ATLAS data. It can also simultaneously fit the purported LEP $ Zb\bar{b}$  excess around the same mass within $1\sigma$ C.L. This neutral Higgs boson is the real component of the right-handed triplet scalar field, whose VEV breaks the $SU(2)_R\times U(1)_{B-L}$ symmetry and gives Majorana masses to the right-handed neutrinos to implement the type-I seesaw. No extra fields need to be introduced to explain the LHC and LEP excesses. This interpretation of the excesses can be {\it completely} tested at the HL-LHC and future Higgs factories, once the precision in the SM Higgs signal strength measurement reaches percent level, which would imply the SM Higgs mixing with an extra scalar below $0.2$ or so. Furthermore, benefiting from the intrinsic structure of the LRSM, the future high-precision data in the $\gamma\gamma$ channel could also be used to determine the $v_R$ scale, or at least provide some information on it, complementary to the direct $W_R$ boson searches. 

\section*{Acknowledgments}

BD thanks Bruce Mellado and Anil Thapa for useful discussions. The work of BD was partly supported by the U.S. Department of Energy under grant No.~DE-SC~0017987. YZ is supported by the National Natural Science Foundation of China under grant No. 12175039 and the Fundamental Research Funds for the Central Universities. BD wishes to acknowledge the Center for Theoretical Underground Physics and Related Areas (CETUP*) and the Institute for Underground Science at SURF for hospitality and for providing a stimulating environment, where part of this work was done. 

\appendix
\section{The decay $h \to H_{3}^0 b\bar{b}$}
\label{sec:appendix}

With $H_3^0$ identified as the 95 GeV scalar, a rare three-body decay of the SM Higgs boson opens up: $h\to H_3^0h^\star \to H_3^0b\bar{b}$. Here we calculate the corresponding partial decay width. The calculation procedure follows Ref.~\cite{threebodydecay}. We skip the messy details and focus only on the order of magnitude estimate. The squared amplitude for the decay $h \to H_{95} b\bar{b}$ is 
\begin{eqnarray}
\label{eqn:ampsq}
\sum |{\cal M}|^2 \simeq \frac{2y_b^2 \mu_{\rm eff}^2 (p_1 \cdot p_2)}{m_h^4} \,,
\end{eqnarray}
where we have neglected the momentum of the SM Higgs propagator as $q^2 \sim (10 \, {\rm GeV})^2 \ll m_h^2$, which is a good approximation. The Yukawa coupling $y_b = m_b/v_{\rm EW} \simeq 0.024$, and the triple scalar coupling $\mu_{\rm eff}$ for the $H_3^0hh$ vertex is 
\begin{eqnarray}
\mu_{\rm eff} = \frac{\sin\theta_1}{2\sqrt2 v_{\rm EW}} (m_h^2 - m_{H_{95}}^2) \simeq 5.3 \, {\rm GeV} 
\left( \frac{\sin\theta_1}{0.4} \right).
\end{eqnarray}
In Eq.~(\ref{eqn:ampsq}) $p_{1,2}$ are respectively the momenta of $b$ and ${\bar b}$ in the final state. Limited by the phase space, its typical value is $p_1 \cdot p_2 \sim (10 \, {\rm GeV})^2$. Then, the partial decay width is given by
\begin{eqnarray}
\Gamma (h \to H_{3}^0 b\bar{b}) &\simeq& \frac{3}{64\pi^3 m_h} \sum |{\cal M}|^2 \nonumber \\ && \times \int_{m_{H_3^0}}^{E_3^{\rm max}} d E_3 \sqrt{E_3^2- m_{H_{3}^0}^2} \,,
\end{eqnarray}
where $E_3$ is the energy of $H_{3}^0$ with its maximal value
\begin{eqnarray}
E_3^{\rm max} = \frac{m_h^2 + m_{H_{3}^0}^2}{2 m_h} \,.
\end{eqnarray}
The integration leads to 
\begin{eqnarray}
\mu_{\rm int}^2 &=& \int_{m_{H_3^0}}^{E_3^{\rm max}} d E_3 \sqrt{E_3^2- m_{H_{3}^0}^2} \nonumber \\
&=& \frac{1}{8m_h^2} 
\left[ m_h^4 - m_{H_{3}^0}^4 + 2 m_h^2 m_{H_{3}^0}^2 \log \left( \frac{m_{H_{3}^0}^2}{m_h^2} \right) \right] \nonumber \\ &\simeq& 60.7 \, {\rm GeV}^2 \,.
\end{eqnarray}
Then the partial width becomes 
\begin{eqnarray}
&& \Gamma (h \to H_{3}^0 b\bar{b}) \simeq \frac{3 y_b^2 \mu_{\rm int}^2 }{32\pi^3 m_h} 
\left( \frac{\mu_{\rm eff}}{m_h} \right)^2 
\left( \frac{p_1 \cdot p_2}{m_h^2} \right) \nonumber \\
&& \simeq 1.4 \times 10^{-11} \, {\rm GeV} \,
\left( \frac{\sin\theta_1}{0.4} \right)^2 
\left( \frac{p_1 \cdot p_2}{(10 \, {\rm GeV})^2} \right) \,. 
\end{eqnarray}
Suppressed by the heavy propagator ($q^2 \ll m_h^2$), phase space, the three-body decay and mixing angle $\sin\theta_1$, the partial decay width for $h \to H_{3}^0 b\bar{b}$ turns our to be much smaller than the SM decay mode  
\begin{eqnarray}
\Gamma (h \to b\bar{b}) &\simeq& \frac{3 y_b^2 m_h}{16\pi} \simeq 0.004 \, {\rm GeV} \, . 
\end{eqnarray}

\bibliographystyle{utcaps_mod}
\bibliography{ref}
  
\end{document}